\documentclass[doublecol]{epl2}

\usepackage{amsmath}
\usepackage{graphicx}
\usepackage{bm}
\usepackage{amssymb}
\usepackage{amsmath}
\usepackage{amssymb}
\usepackage{booktabs}
\usepackage{multirow}
\usepackage{array}

\title{Fabry-P\'{e}rot interference in 2D low-density Rashba gas}
\shorttitle{Fabry-P\'{e}rot interference in 2D low-density Rashba gas} 

\author{Yuan-Qiao Li\inst{1,\footnote{These authors contributed equally to this work}} \and Xi-Rong Chen\inst{2,\footnote{These authors contributed equally to this work}} \and Wei Luo\inst{2,3,\footnote{Corresponding author: jxluow@163.com}} \and Tao Zhou\inst{4}\and Wei Chen\inst{2,5,\footnote{Corresponding author: pchenweis@gmail.com}}}
\shortauthor{Yuan-Qiao Li \etal}

\institute{
  \inst{1} College of Science, Nanjing University of Aeronautics and Astronautics, Nanjing 210016, China\\
  \inst{2} National Laboratory of Solid State Microstructures and School of Physics, Nanjing University, Nanjing, 210093, China\\
  \inst{3} School of Science, Jiangxi University of Science and Technology, Ganzhou 341000, China\\
  \inst{4} Guangdong Provincial Key Laboratory of Quantum Engineering and Quantum Materials, GPETR Center for Quantum Precision
Measurement, SPTE, South China Normal University, Guangzhou 510006, China\\
  \inst{5} Collaborative Innovation Center of Advanced Microstructures, Nanjing University, Nanjing 210093, China
  }

\abstract{In mesoscopic electronic systems, the Fabry-P\'{e}rot (FP) oscillation is
observed in various 1D devices. As for higher dimensions, numerous transverse
channels usually lead to dephasing
that quenches the overall oscillation of the conductance.
Up to now, the FP oscillation
in 2D electronic systems is only reported in
graphene-based devices and very recently, the \emph{pn}
junctions of inverted InAs/GaSb double quantum well [Phys. Rev. X 10, 031007 (2020)]. In the latter,
the band shape of a sombrero hat plays an essential role, which
introduces a novel mechanism of electron-hole
hybridization for the 2D FP oscillation.
In this work, we propose that such a scenario can be
generalized to the 2D planar junction composed of
low-density Rashba gas, where the
band bottom possesses sombrero hat shape as well.
We show that the backscattering between the outer and inner
Fermi circles dominates the FP interference and significantly
suppresses the dephasing effect between different transverse channels,
which leads to a visible oscillation
of the tunneling conductance. Specially,  the visibility of the oscillating pattern can be enhanced
by applying interface barriers, which in contrast to that in the InAs/GaSb double quantum well.
Our results provide a promising way for the implementation of
the FP oscillation in the 2D electron gas.}

\begin{document}
\maketitle

\section{Introduction}

Fabry-P\'{e}rot (FP) interference occurs between a pair of reflectors,
where a wave propagating in between can undergo multiple reflections
with different coherent paths interfering with each other.
In optics, it is manifested as the fine structure of the interference patterns
of the transmissivity as the distance between reflection surfaces varies.
As a universal property of coherent waves,
FP oscillation generally exists in various physical systems.
In mesoscopic systems for example, the coherent electronic wave
also exhibits FP interference, which usually manifests as
the conductance oscillation with varying energy.
The FP oscillation has been observed in several
1D systems, such as carbon nanotubes~\cite{Nature665}
and quantum Hall edge states~\cite{PRL2523,Nature415,PNAS5276,PRL206806,NaturePhysics563}.
However, it is quite challenging to realize
such a novel effect in higher dimensions,
because the existence of numerous transverse
channels with different oscillating phases
can easily quench the overall signature.
One exception is the 2D \emph{pnp} (or \emph{npn}) junctions made of single-layer~\cite{NaturePhysics222,NatureCommunications2342,APL223102,PRB121414}
or multi-layer graphene~\cite{NatureCommunications1239,PRL116601},
in which the Klein or anti-Klein tunneling together with the unique
Dirac dispersion enable the survival of FP oscillation
after averaging over all transverse channels.

Most recently, FP oscillation was observed in another 2D system,
the \emph{pnp} junctions
of the inverted InAs/GaSb double quantum well~\cite{PRX031007}
which is well known for its nontrivial band topology~\cite{PRL236601,PRB235311,PRL096802,NatureNanotechnology593}. Its band structure is featured by the
band inversion and electron-hole hybridization~\cite{PRB842,PRL3034,PRL4613,PRB11915}.
Different from a conventional 2D band structure,
such a band shape resembles a sombrero hat,
which is key to the observation of 2D FP oscillation~\cite{PRX031007}.
Specifically, the electron (\emph{e}) and hole (\emph{h}) states coexist
close to the band bottom and
the \emph{e-h} reflection dominates the FP
interference. It significantly reduces the phase difference
of oscillation between the transmission functions
of different transverse channels and gives
rise to visible oscillation of the conductance,
in contrast to the conventional \emph{e-e} reflection.
This discovery expanded the field of electron optics to
include materials that exhibit band inversion and hybridization.
Notably, the band shape similar to the sombrero hat
can be found in various condensed matter systems, of which
the typical example is the 2D electron gas with Rashba
spin-orbit coupling~\cite{Jetplett78} [see fig.~\ref{fig1}].
Then a natural question that arises
is whether the FP oscillation can be observed in
these systems as well.

In this work, we give an affirmative answer to this question
by studying the FP interference in
the 2D junctions of the low-density
Rashba gas.
Specifically, for the normal metal-Rashba gas-normal metal (NRN)
junctions, reflections that take place between the outer and inner
Fermi circles induce visible FP oscillation in
the total transmission function,
which can be probed by the oscillation of the differential conductance.
Moreover, we found that the interface barriers can enhance the
visibility of the oscillation pattern
due to the helical spin texture
of the Rashba gas.
Our findings pave the way towards the implementation of 2D
FP interference in a new electronic system.

This paper is organized as follows: We
study the scattering problem in
in the NRN junctions and analyze the
FP oscillation of the conductance in the second section.
The effect due to the magnetic field is investigated
in the third section.
Finally, the discussion on the experimental implementation
of our proposal and
a brief summary are given in the fourth section.

\section{FP interference in NRN junctions}

\begin{figure}
\centering
\includegraphics[width=0.45\textwidth]{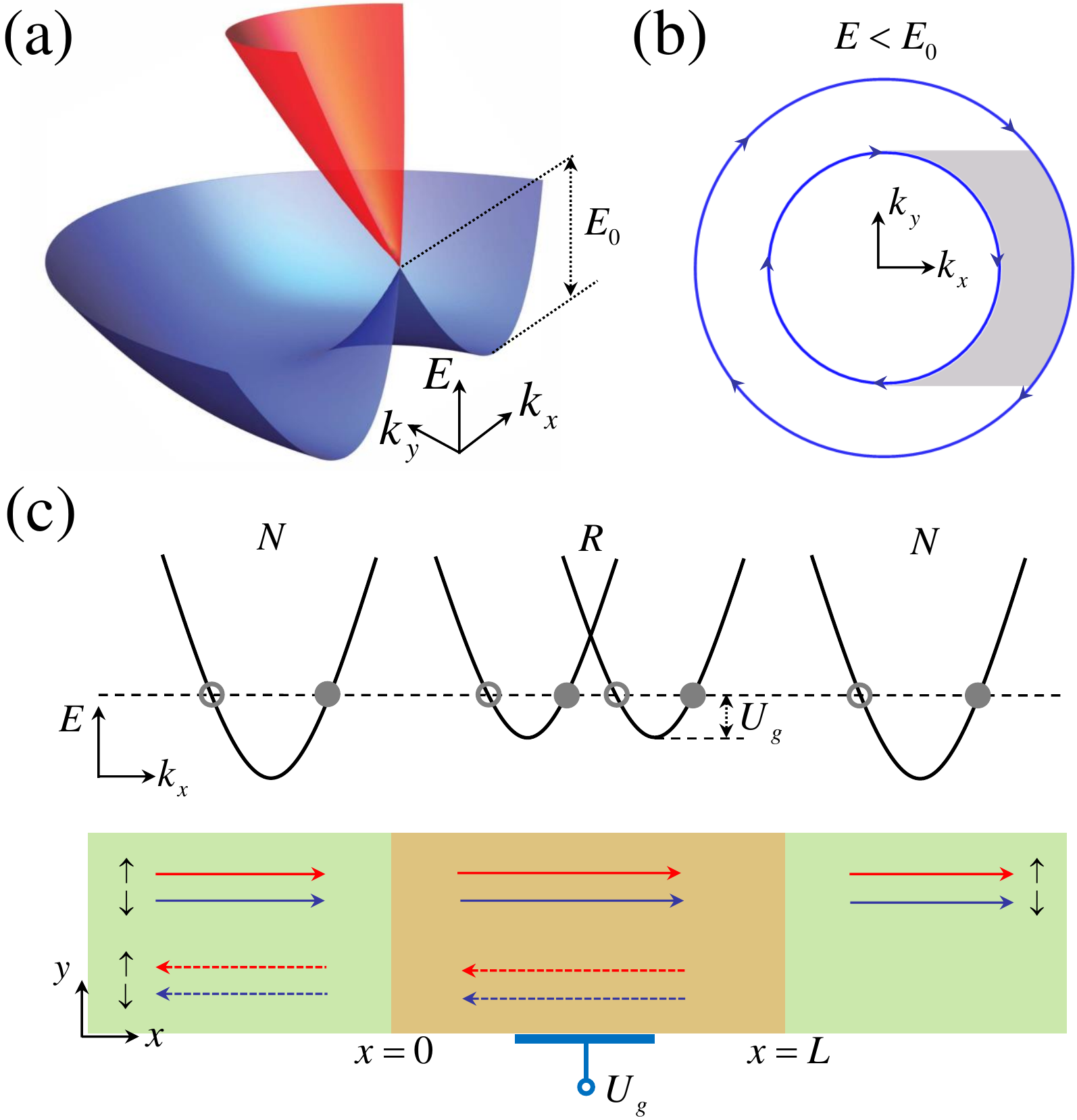}
\caption{(Color online). (a) Three-dimensional plot of Rashba energy band. (b) The doubly connected
Fermi surface below the Dirac point, blue arrows indicate the
directions of electron spin. (c) Schematic illustration of the reflection and
transmission processes in the junction, the red and blue arrows denote the spin
up and spin down electron states, respectively.}
\label{fig1}
\end{figure}

We consider the 2D NRN junctions lying in the $x$-$y$ plane
which consist of
the low-density Rashba gas sandwiched by two pieces
of normal metals with spin degeneracy; see fig.~\ref{fig1}(c).
The whole system can be captured
by the effective model as
\begin{equation}
\label{HZ}
\begin{split}
&H \!=\! {H_N}\big[\Theta(-x)+\Theta(x-L)\big] \!+\! {H_R}\Theta(x)\Theta(L-x) \!+\! U(x),\\
&{H_N} = \frac{\hbar^2\bm{k}^2}{2m_0} - \mu,\,
{H_R} = \frac{\hbar^2\bm{k}^2}{2m_1} - {U_g}
+ \lambda_R (k_x\sigma_y-k_y\sigma_x),
\end{split}
\end{equation}
where $H_N$ and $H_R$ describe
the normal metal and the Rashba gas, respectively,
with $\Theta(x)$ the Heaviside
unit step function,
$\bm{k}=(k_x, k_y)$
is the momentum
and $m_{0}$ and $m_1$ are the effective masses.
The energy in the normal metal is measured from the chemical potential $\mu$.
In the middle region, a gate voltage $U_g$ is applied
to adjust the energy of the bands, $\sigma_{x,y}$ are Pauli matrices
for electronic spin and $\lambda_R$ is the Rashba coefficient, which can reach
as high as
2$\mathrm{eV\, {\AA}}$ in the experiment~\cite{PRB081407}.
The interface barriers are simulated by two
$\delta$-function potentials $U(x)=U_1\delta(x)+U_2\delta(x-L)$
with strengths $U_{1,2}$.

Solving $H_R$ yields two bands ${E_ \pm }(k) = {\hbar^2( k\pm{k_0}})^2/(2m_1)- E_0 - {U_g}$
with $k_0=m_1\lambda_R/\hbar^2$ and $E_0=\hbar^2k_0^2/(2m_1)$.
The bands exhibit a spin dependent splitting as shown in fig.~\ref{fig1}(a).
Notably, the band bottom of the Rashba gas resembles that of the inverted InAs/GaSb
quantum well \cite{PRX031007}, which is key to the 2D FP interference.
Here, we focus on the low-density limit of the Rashba gas
such that the Fermi energy only intersects the $E_-(k)$ branch.
In the following, we refer the inner and the outer
Fermi circles as \emph{h} and \emph{e}, respectively,
by noting that the velocity and the momentum
have opposite sign for the inner Fermi circle [cf. figs.~\ref{fig1}(a) and ~\ref{fig1}(b)].

\begin{figure*}
\centering
\includegraphics[width=0.85\textwidth]{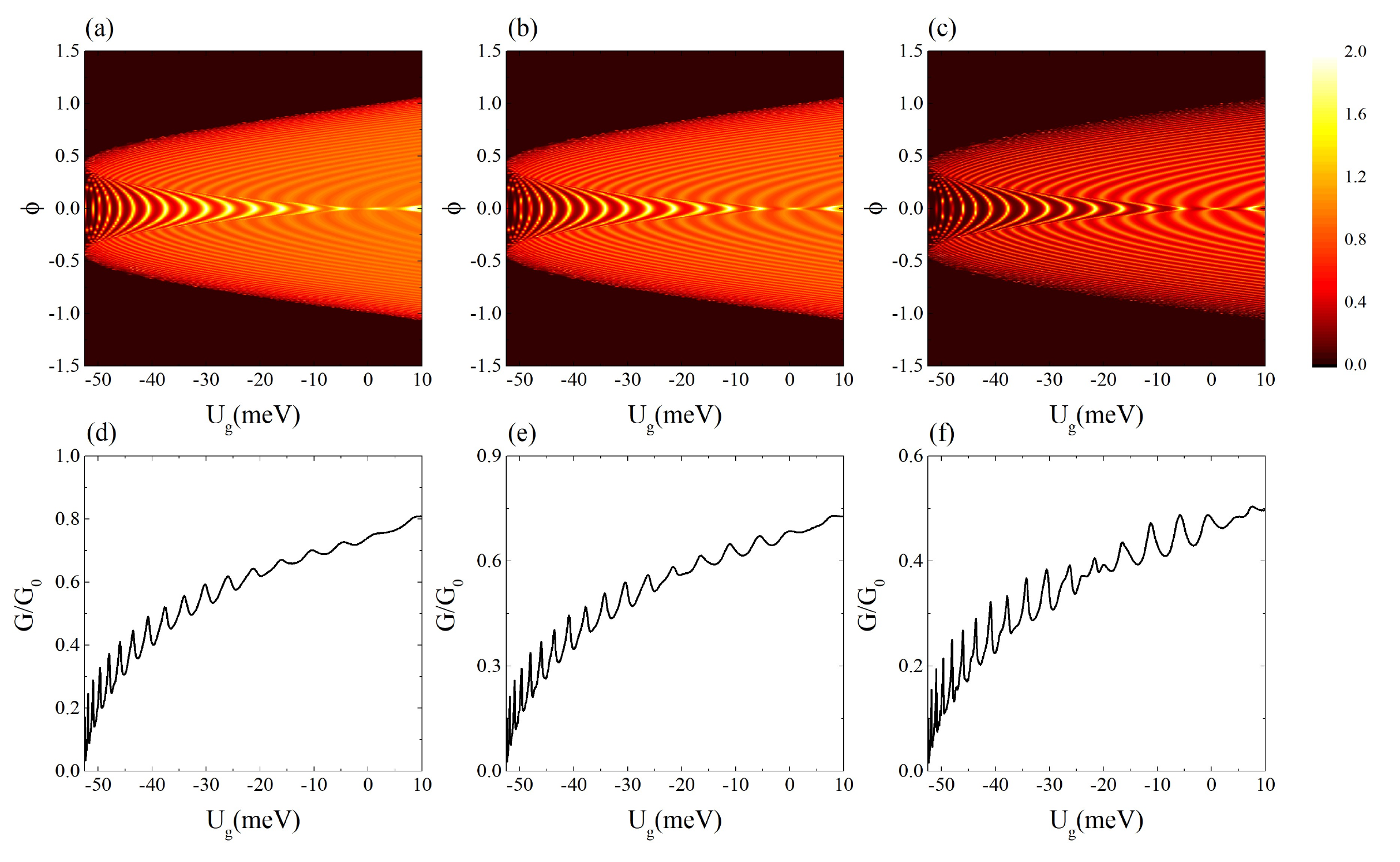}
\caption{(Color online). Top rows illustrate the numerical results of the total
transmission coefficient $T$ as a function of $U_g$ and $\phi$, and bottom rows
exhibit the numerical results of the conductance $G/G_0$. The interface barriers
in the left, middle, and right panels are $Z_1=Z_2=0,0.5$ and $1$, respectively.
The other parameters are set as $m_1=m_0=0.2m_{e}$ with $m_{e}$ the free-electron
mass, $L=100\mathrm{nm}$, $\lambda_{R}=0.2\mathrm{eV}\cdot \mathrm{nm}$ and
$E_F=0.3\mathrm{eV}$.}
\label{fig2}
\end{figure*}

Next we solve the spin resolved scattering problem in the NRN junction.
In the normal metal, no spin-orbit coupling exists
so the energy bands are spin degenerate.
In the middle region, the band splitting introduces
two branches of scattering states in the Rashba gas; see fig.~\ref{fig1}(c).
Two Fermi circles have the same/opposite winding direction of the spin as
the Fermi energy
lies below/above the central band touching point.
Suppose that the size of
the junction along the $y$ direction is much larger than the Fermi wave length
so that $k_y$ is conserved during scattering. For
an electron with a spin $\sigma(=\uparrow,\downarrow)$ and an energy $E$ incident from
$x=-\infty$, the wave functions in the left, the middle and the right
regions are
\begin{equation}
\begin{split}
\psi_L^\sigma&=\Big[|\sigma\rangle e^{ik_xx}+\sum_{\sigma'=\uparrow,\downarrow}r^\sigma_{\sigma'}|\sigma'\rangle e^{-ik_xx}\Big] e^{ik_yy},\\
\psi_M^\sigma&=\sum_{\alpha,\eta}s_\eta^\alpha(u_\eta^\alpha, v_\eta^\alpha)^{\text{T}}e^{i\eta k^{\alpha}_{x}x+ik_yy},\\
\psi_R^\sigma&=\sum_{\sigma'=\uparrow,\downarrow}t^\sigma_{\sigma'}|\sigma'\rangle e^{ik_xx+ik_yy},
\end{split}
\end{equation}
where $r_{\sigma'}^\sigma$ and $t_{\sigma'}^\sigma$ are the amplitudes of the reflected
and transmitted waves with spin $\sigma'$, respectively.
The coefficients $s^\alpha_\eta$ with $\alpha=1, 2$ and $\eta=\pm$
correspond to four scattering waves in the middle region.
The spinor parts of the scattering waves in the Rashba region
is defined as $u_{\eta}^{\alpha}=(-1)^{\alpha}(k_{y}+i\eta k_{x}^{\alpha})$ and
$v_{\eta}^{\alpha}=\sqrt{k_{y}^{2}+{k_{x}^{\alpha}}^{2}}$.
The $x$-direction momentums in different regions are
$k_x=\sqrt{2m_0(E+\mu)/\hbar^{2}-k_y^2}$ and
$k^\alpha_x=\sqrt{(k_0+(-1)^{\alpha}\sqrt{2m_1(U_g+E_{0}+E)/\hbar^{2}})^2-k_y^2}$,
respectively.

All the scattering coefficients can be solved by the boundary conditions
at two interfaces as
\begin{equation}
\begin{split}
&\psi_L^\sigma(0)=\psi_M^\sigma(0)\equiv\psi_0^\sigma, \ \
\psi_M^\sigma(L)=\psi_R^\sigma(L)\equiv\psi_L^\sigma,\\
&\partial_x\psi_M^\sigma(0)-\frac{m_1}{m_0}\partial_x\psi_L^\sigma(0)=(2k_F Z_1-ik_0\sigma_y)\psi_0^\sigma,\\
&\frac{m_1}{m_0}\partial_x\psi_R^\sigma(L)-\partial_x\psi^\sigma_M(L)=(2k_FZ_2+ik_0\sigma_y)\psi_L^\sigma,
\end{split}
\end{equation}
where $Z_{j=1,2} =m_1U_j/(\hbar^{2}k_F)$ are the reduced barrier strengths
defined by the Fermi wave vector $k_F=\sqrt{2m_0\mu/\hbar^{2}}$.
The total reflection and transmission probabilities are obtained as
$R^\sigma=\sum_{\sigma'}|r^\sigma_{\sigma'}|^2$ and $T^\sigma=\sum_{\sigma'}|t^\sigma_{\sigma'}|^2$
from which the current conservation condition $R^\sigma+T^\sigma=1$
for each transverse channel $k_y$ can be verified.

The FP interference for a 2D junction can be revealed
by the oscillation pattern of the transmission
$T(\phi,U_g)=\sum_\sigma T^\sigma$ as a function of the
gate voltage $U_g$ in various
transverse modes labeled by the incident angle
$\phi=\tan^{-1}(k_y/k_x)$; see figs.~\ref{fig2}(a)-(c).
The oscillation patterns share the same feature as those
in the InAs/GaSb double quantum well~\cite{PRX031007}. It contains
two typical regions marked with
the bright central stripes and the darker outlying ones
indicating the maximum transmissions.
The key observation is that the transmission functions for energies
close to the band bottom exhibit a weak $\phi$-dependence
which can be tracked by the bright stripes in figs.~\ref{fig2}(a)-(c).
This part of interference pattern corresponds to the
\emph{e-h} and \emph{h-e} reflections between the outer and inner
Fermi circles as shown in fig.~\ref{fig1}(a),
the same as that the InAs/GaSb double quantum well~\cite{PRX031007}.
It results in a weak dependence of the momentum transfer
$\Delta k_x$ and accordingly, the accumulated FP phase
on the transverse momentum $k_y$,
which facilitates the 2D FP oscillation~\cite{PRX031007}.
On the contrary, for the \emph{e-e} and \emph{h-h} reflections that occur
within the inner or outer Fermi circles for higher energies,
$\Delta k_x$ varies significantly with $k_y$
which only contribute a smooth background of the conductance.

\begin{figure*}
\centering
\includegraphics[width=0.85\textwidth]{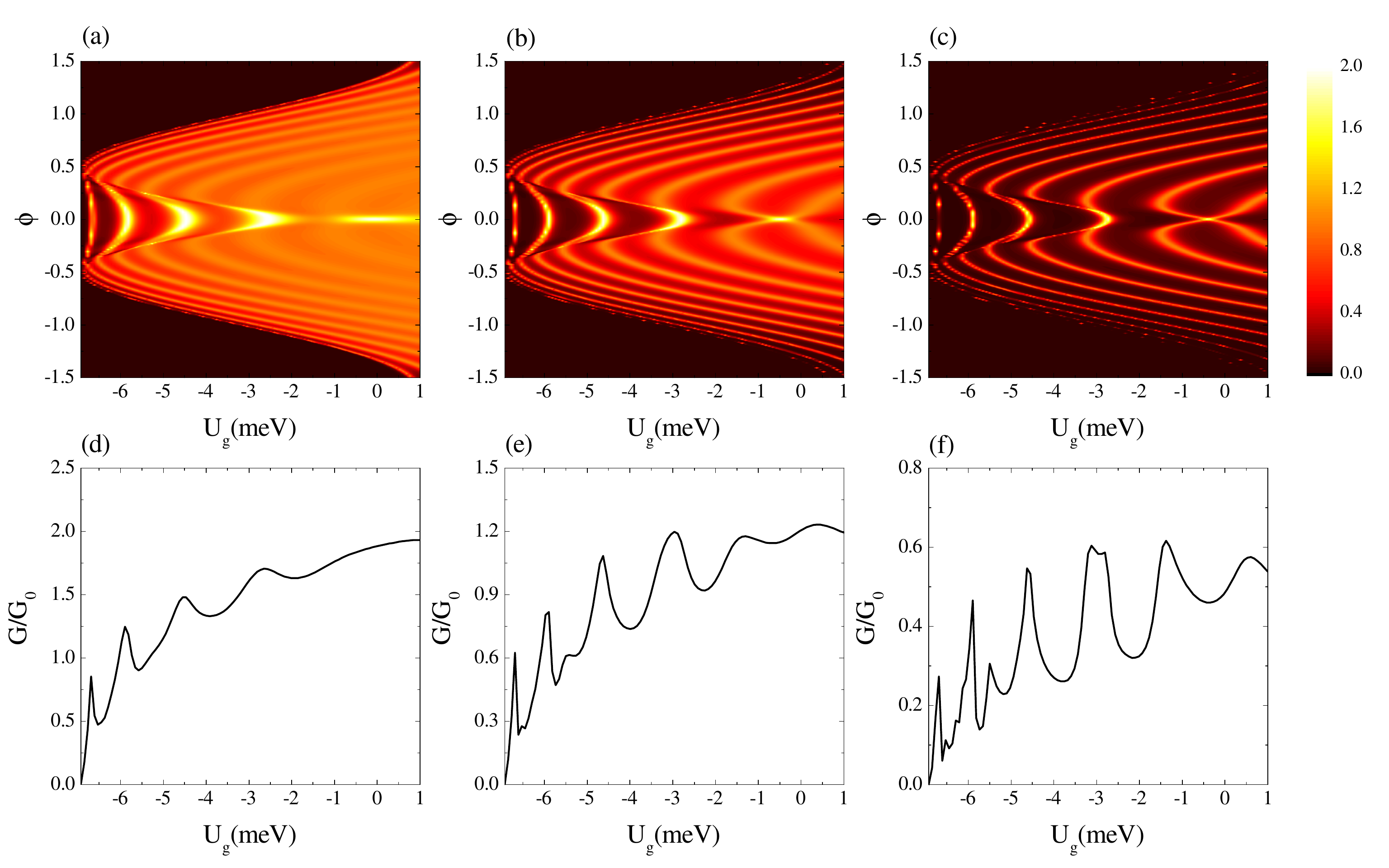}
\caption{(Color online). Top rows illustrate the numerical results of the total
transmission coefficient $T$ as a function of $U_g$ and $\phi$, and bottom rows
exhibit the numerical results of the conductance $G/G_0$. The interface barriers
in the left, middle, and right panels are $Z_1=Z_2=0,0.5$ and $1$, respectively.
The other parameters are set as $m_1=m_0=0.2m_{e}$ with $m_{e}$ the free-electron
mass, $L=300\mathrm{nm}$, $\lambda_{R}=0.02\mathrm{eV}\cdot \mathrm{nm}$ and
$E_F=2.2\mathrm{meV}$.}
\label{fig3}
\end{figure*}

The weak dependence of $\Delta k_x$ on $k_y$ indicates that
a visible oscillation will
remain after summing up the
transmission probabilities
over all transverse channels,
which can be probed by the differential conductance
\begin{equation}
G(U_g) = {G_0}\int_{0}^{\phi_c} T(\phi,U_g)\cos \phi d\phi,
\end{equation}
where ${G_0} = \frac{{{e^2}}}{h}\frac{{{W}{k_F}}}{{\pi}}$ is the conductance
for the normal metal stripe with a width $W$. The incident energy
is set to $E=0$ and the integral interval is bounded
by the critical incident angle $\phi_c=\sin^{-1}[(k_0+\sqrt{2m_{1}(E_{0}+U_{g})/\hbar^{2}})/k_F]$. The numerical results
of $G$ as a function of the gate voltage $U_g$ are shown in figs.~\ref{fig2}(d)-(f),
which exhibit a visible oscillation of the conductance below the
band crossing energy of the Rashba gas ($U_g=0$meV).

Apart from the similarity to the InAs/GaSb quantum well,
the Rashba gas has its unique feature. Note that two
Fermi circles below the Dirac point possess the same
spin winding [cf. fig.~\ref{fig1}(a)], so that the \emph{e-e} (\emph{h-h})
reflection is strongly suppressed for small incident angle $\phi$
due to the opposite spin polarization of the incident and reflected states.
As a result, the \emph{e-h} (\emph{h-e}) reflection dominates the FP interference
with a nearly conserved spin.
Such a scenario is expected to be robust against interface barriers.
In figs.~\ref{fig2}(a)-~\ref{fig2}(c), one can see that
the interface barriers make the transmission stripes even sharper,
which indicates that the \emph{e-h} scattering
is enhanced by the interface barrier while the opposite
spin orientation for the \emph{e-e} (\emph{h-h})
scattering still prohibit its occurrence.
Accordingly, the visibility of the
conductance oscillation increases as well
despite a reduction of its magnitude; see figs.~\ref{fig2}(d)-~\ref{fig2}(f).

In the above calculation, we use a large value for $\lambda_R$, i.e., $0.2\mathrm{eV}\cdot \mathrm{nm}$, which was experimentally observed in a $Bi_xPb_{1-x}/Ag(111)$ surface alloy~\cite{PRB081407}. It is more promising that an experiment to observe this effect would be based on some sort of semiconductor heterostructure, which would feature $\lambda_R$ values about an order of magnitude lower (i.e. $0.02\mathrm{eV}\cdot \mathrm{nm}$). The most promising materials and relative parameters are listed in table ~\ref{tab1}, where the Rashba energy of split states $E_0$ has been computed from the experimentally accessible quantities $m_1$ and $\lambda_R$. Thus we calculate the transmission coefficient and conductance with $\lambda_{R}=0.02\mathrm{eV}\cdot \mathrm{nm}$ as shown in fig.~\ref{fig3}. As one can see, the conductance spectra exhibit a distinct oscillation, which can be enhanced by the interface barriers. The main effect due to a smaller Rashba strength is a reduction of the number of oscillating peaks as shown in figs.~\ref{fig3}(d)-~\ref{fig3}(f). We conclude that the FP interference and corresponding conductance oscillations can still be observed for values of $\lambda_{R}$ about an order of magnitude lower.

\begin{table}
\caption{Selected materials and parameters characterizing the spin splitting: effective masses $m_1$, Rashba coefficient $\lambda_R$, and Rashba energy of split states $E_0$.}
\centering
\label{tab1}
\setlength{\tabcolsep}{1.2mm}{
\begin{tabular}{cccc}
\toprule
\multirow{2}{*}{Material} & $m_{1}$ & $\lambda_R$ & $E_{0}$ \\
 & ($m_{e}$) & ($\mathrm{eV}\cdot\mathrm{nm}$) & ($\mathrm{meV}$)\\

\hline

$\mathrm{PtSe_{2}/MoSe_{2}}$ & \multirow{2}{*}{0.81} & \multirow{2}{*}{0.13} & \multirow{2}{*}{89.8}  \\
 heterostructures\cite{APL203501} &&& \\

\specialrule{0em}{3pt}{3pt}

$\mathrm{In_{x}Ga_{1-x}As}$ heterostructures & \multirow{2}{*}{0.25} & \multirow{2}{*}{0.037} & \multirow{2}{*}{2.25}  \\
along Au(111) surface\cite{PSS1779} &&& \\

\specialrule{0em}{3pt}{3pt}

Te-terminated & \multirow{2}{*}{0.1489} & \multirow{2}{*}{0.35} &  \multirow{2}{*}{119.7} \\
 surface of BiTeI\cite{Jetp437} &&& \\

\bottomrule
\end{tabular}}
\end{table}

\section{magnetic field effect}

We now discuss the effect of the magnetic field $B$ on the FP oscillations,
which is usually investigated simultaneously
in the experiment of 2D electronic FP oscillation.
Since the transverse momentum $k_y$ is approximately conserved during scattering,
it can be chosen as a parameter. In this way, the 2D system is decomposed
into a set of 1D channels labeled by $k_y$. The
Landau gauge $\bm{A}=(0, Bx)$ is adopted so that
the Peierls substitution $\bm{k}\rightarrow \bm{k}-e\bm{A}/\hbar$ retains
the conservation of $k_y$. We perform numerical simulation
on a square lattice through the substitutions of $k_{x,y}\rightarrow a^{-1}\sin k_{x,y}a$
and $k_{x,y}^2\rightarrow 2a^{-2}(1-\cos k_{x,y}a)$, with $a$ the
fictitious lattice constant.
Keeping $k_y$ a parameter and performing Fourier transformation yields the
Rashba Hamiltonian on the discrete lattice as
\begin{equation}
\mathcal{H}_R(k_y)=\sum_{i} c_i^{\dagger}H_{ii}c_i+\sum_{i}(c_i^{\dagger}H_{i,i+a}c_{i+a}+\text{H.c.}),
\end{equation}
where $c_i=(c_{i,\uparrow},c_{i,\downarrow})$ is the Fermi operator on site $i$ with both spin components
and
$H_{ii}$ and $H_{i,i+a}$ are $2\times 2$ block matrices with the explicit form of
\begin{equation}
\begin{split}
&H_{ii}=\frac{\hbar^2}{m_1a^2}\left(2-\cos k_ya\right)-U_g-\frac{\lambda_R}{a}\sin k_ya\sigma_x,\\
&H_{i,i+a}=-\frac{\hbar^2}{2m_1a^2}+\frac{\lambda_Ri}{2a}\sigma_y,\\
\end{split}
\end{equation}
The lattice model for the normal metal can be obtained in a similar way, which writes
\begin{equation}
\mathcal{H}_N(k_y)=\sum_{j} C_jd_j^{\dagger}d_j-\frac{\hbar^2}{2m_0a^2}\sum_{j}(d_j^{\dagger}d_{j+a}+\text{H.c.}),
\end{equation}
where $d_j=(d_{j,\uparrow},d_{j,\downarrow})$ denotes the spinful fermion operator in the normal metal, and $C_j=\hbar^2(2-\cos k_ya)/(m_0a^2)-\mu$.

\begin{figure}
\centering
\includegraphics[width=0.48\textwidth]{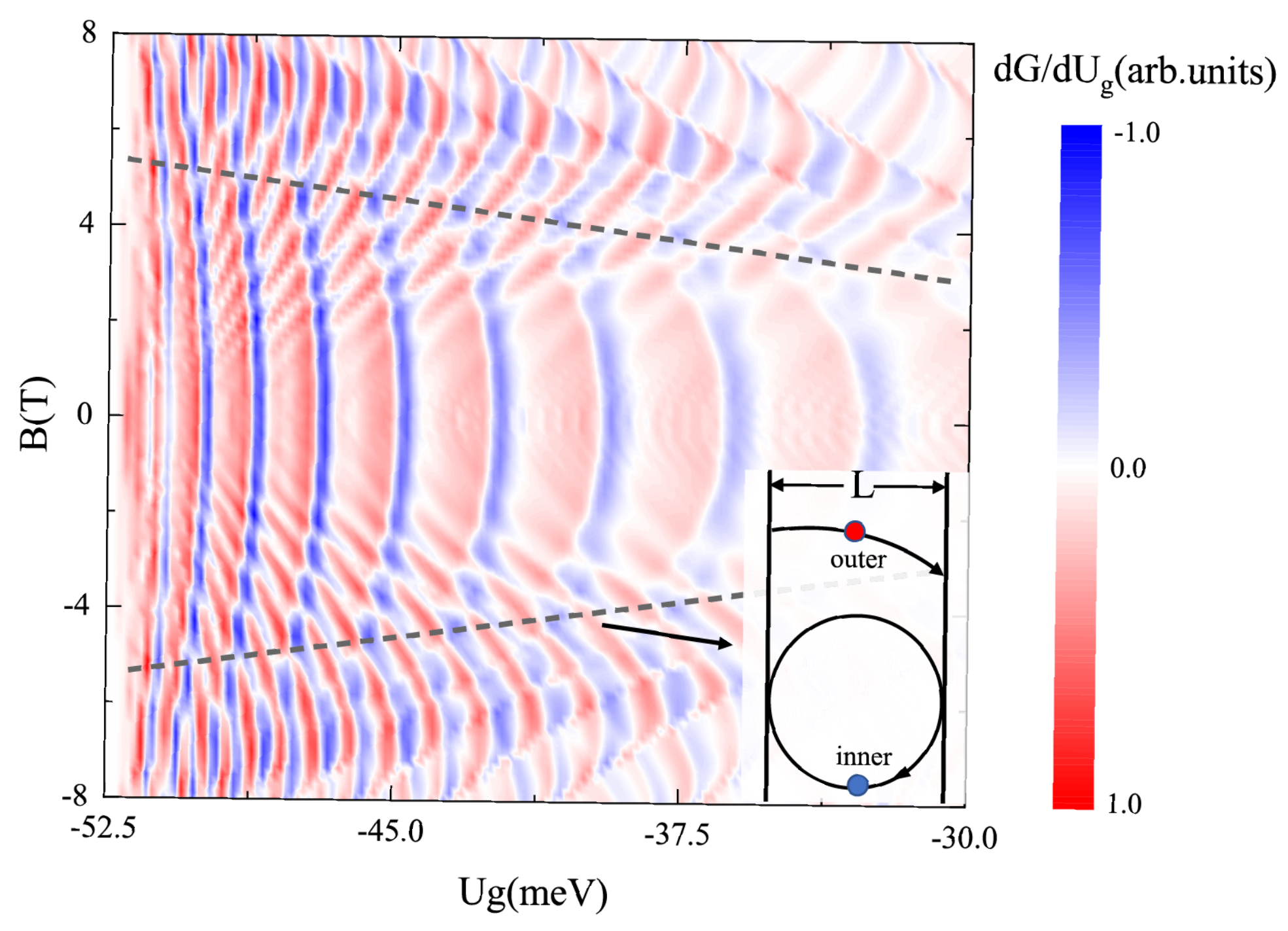}
\caption{(Color online). Calculated dependence of $dG/dU_g$ on $U_g$ and $B$ in the case of $Z=0$. The inset shows schematic trajectories for inner
and outer Fermi circle states at the critical magnetic field $B_c$ of the inner states, which is represented by the dotted lines.
The other parameters are the same as those in fig.\ref{fig2}.}
\label{fig4}
\end{figure}

For a given $k_y$, the transmission probability is calculated using the KWANT package~\cite{NJP063065}.
A summation of the transmissions of all $k_y$ channels yields the conductance
of the 2D junction. We plot the modulation of the
oscillation pattern of the differential conductance $dG/dU_g$ induced by the magnetic field in fig.\ref{fig4}.
For a small $B$, the conductance pattern exhibits a weak $B$-dependence especially
for the energy around the band bottom.
As $B$ increases, the tripes tracking the resonance bend toward lower $U_g$.
Comparing $dG/dU_g$ at $B=0$ and that at $B\simeq4\,\rm T$, there is an interchange
of the minima and maxima except for certain additional structures, which indicates
a phase shift of $\pi$ in the oscillation pattern of the conductance.
As $B$ increases further, the oscillation pattern tends to
be less visible accompanied by a reversal of the bending direction
of the resonant stripes. The two different regimes
correspond to \emph{e-h} reflection and \emph{e-e} (\emph{h-h})
reflection, respectively, and a boundary between them
is marked by the dashed lines in fig.\ref{fig4}.
In the semiclassical picture, the cyclotron motion of
an electron possesses a radius of $r=\hbar k_F/eB$, with the Fermi wave vector $k_F$.
The holes (\emph{h}) in the inner Fermi surface possess a smaller
cyclotron radius (cf. fig.~\ref{fig1}), which decreases with increasing $B$.
Once it decreases to the threshold $r_c=L/2$ at
a critical field $B_c$, where $L$ is the length of the Rashba region [cf. fig.~\ref{fig1}(c)],
the electron in the inner Fermi surface cannot transmit through the Rashba region,
see the inset of fig.~\ref{fig4}.
As a result, the \emph{e}-\emph{h} reflection is suppressed,
which quenches the FP oscillation of the conductance.
The critical magnetic field $B_c$ as a function of the gate
voltage $U_g$ is plot by the dashed lines in fig.~\ref{fig4}.

\section{Discussions and summary}

We would like to discuss the experimental implementation of our proposal.
The 2D low-density Rashba gas is a crucial building block in our proposal,
which has been achieved in the heterostructures~\cite{PRL1335,Springer191,PRL256601}.
These systems are fabricated by the band engineering techniques
taking advantage of the interfacial/low-dimensional effects,
in which the Rashba coefficient can be precisely
controlled by an electric field. Moreover, the physical
parameters such as the chemical potential in such
heterostructures can be easily tuned by a gate voltage.
For example, it was reported that the Rashba energy and the
Fermi energy in a two-dimensional electron gas can be
tuned by a controlled change of stoichiometry in an
artificial surface alloy~\cite{PRB081407}, whose Rashba
parameter $\alpha_R$ can reach several $\rm eV\,\AA$.
These progresses pave a way to realize our proposal in the NRN junctions.

In summary, we propose to realize the FP interference in the 2D low-density Rashba gas. The low-density Rashba gas exhibits a unique spin texture, resulting in a small probability of reflection within the inner and outer Fermi circles, which favors reflections between the inner and outer Fermi circles that determine the FP interference. For a spin-orbit coupled system, a larger Rashba enengy of split states has a more profound effect on the FP interference. According to the results from our models, there is a FP oscillation of the conductance versus gate voltage $U_g$ in the 2DEG NRN junction,
which benefits from the interface barriers. With increasing the interface barriers, the reflections between
 the outer and inner Fermi circles are further enhanced, and thus the FP interference becomes more visible. Therefore,
 the resolution of the FP interference in response to the interface barriers can be an observable to detect the unique
 spin textures of the low-density rashba gas. Moreover, the dependence of magnetic
field is included. The lines of constant transmission bend toward
lower $U_g$ with increasing the magnetic field, and a phase shift
of $\pi$ occurs at certain $B$. Further increasing the magnetic
field will suppress the interference phenomenon. Based on the
above results, we perfected the field of electron optics
to include materials that exhibit band hybridization.

\acknowledgments
We thank Oded Zilberberg and Antonio \v{S}trkalj for helpful discussions.
This work was supported by the National
Natural Science Foundation of
China under Grant No. 12074172 (W.C.), the startup
grant at Nanjing University (W.C.)
and the Excellent Programme at Nanjing University.

\end{document}